\documentclass[a4paper,12pt]{article}
\usepackage{cite}
\usepackage{amssymb}
\usepackage{amsmath}
\usepackage{amsfonts}
\usepackage{amsthm}
\usepackage{mathtools}
\usepackage[utf8]{inputenc}
\usepackage{color}
\usepackage{graphicx}
\usepackage{color}
\usepackage[active]{srcltx}

\usepackage{here}
\usepackage{cite}
\usepackage{soul}
\usepackage[affil-it]{authblk}

\input paperdef

\begin{document}

\thispagestyle{empty}

\title{
$N=1$ trinification from dimensional reduction of $N=1$, $10D$ $E_8$ over $SU(3)/U(1)\times U(1)\times Z_3$ \\ and its phenomenological consequences}
\date{}
\author{ \hspace*{7mm} George Manolakos$^1$\thanks{email: gmanol@central.ntua.gr}~, 
Gregory Patellis$^1$\thanks{email: patellis@central.ntua.gr}~
and George Zoupanos$^{1,2,3,4}$\thanks{email: George.Zoupanos@cern.ch}\\
{\small
$^1$ Physics Department,   National Technical University, 157 80 Zografou, Athens, Greece\\
$^2$ Max-Planck Institut f\"ur Physik, F\"ohringer Ring 6, D-80805
  M\"unchen, Germany \\
 $^3$ Max-Planck-Institut f\"ur Gravitationsphysik (Albert-Einstein-Institut), Potsdam, Germany\\
$^4$ Theoretical Physics Department, CERN, Geneva, Switzerland \\  
}
}

\maketitle

\begin{abstract}
We present an extension  of the Standard Model that results from the dimensional reduction of the $\mathcal{N}=1$, $10D$ $E_8$ group over a $M_4 \times B_0/ \mathbf{Z}_3 $ space, where $B_0$ is the nearly-K\"ahler manifold
$SU(3)/U(1) \times U(1)$ and $\mathbf{Z}_3$ is a freely acting discrete group on $B_0$. Using the Wilson flux breaking mechanism
we are left in four dimensions with an $\mathcal{N}=1$ $SU(3)^3$ gauge theory. Below the unification scale we have a two Higgs doublet model in a split-like supersymmetric version of the Standard Model, which yields third generation quark and light Higgs masses within the experimental limits and predicts the LSP $\sim 1500~GeV$.
\end{abstract}


\section{Introduction}\label{intro}
The origins~of our study lie in context~of the pioneering work of Forgacs-Manton (F-M) and Scherk-Schwartz~(S-S) who studied the~Coset Space Dimensional Reduction~(CSDR) \cite{Forgacs:1979zs,Kapetanakis:1992hf, Kubyshin:1989vd} and the group~manifold reduction \cite{Scherk:1979zr}, respectively. Roughly,~these mechanisms share a common~view with another (almost) contemporary~framework to them, that~of the superstring theories \cite{Green:1987sp}, in the~sense that they result in GUTs which~originate from a spacetime that~is extra-dimensional, as, in particular,~in the heterotic string \cite{Gross:1985fr}. The two~approaches found contact in the sense~that CSDR incorporated the~predictions of the heterotic~string, that is the number of extra~dimensions and the gauge group of~the initial theory. In both,~S-S and F-M, mechanisms, in~the higher-dimensional theory, the~gauge and scalar sector are unified and,~specifically in the CSDR case,~fermions of the higher-dimensional~theory lead to Yukawa interactions~in the $4D$ one. Also, it is remarkable~that it can lead to $4D$ chiral~theories  \cite{Manton:1981es}. An additional~important property of the CSDR breaks the~original supersymmetry of a theory,~either completely when is~reduced over symmetric cosets or~softly in the case of the~$6D$ non-symmetric \cite{Manousselis:2001xb} ones which~are all nearly-K\"ahler manifolds~admitting a connection with~torsion \cite{Chatzistavrakidis:2008ii,LopesCardoso:2002vpf,Chatzistavrakidis:2009mh} (see also \cite{Klaput:2011mz}).     

Performing the dimensional~reduction of an $\mathcal{N}=1$ supersymmetric~gauge theory, a very important~and desired property is the amount of~supersymmetry of the initial~theory to be preserved in~the $4D$ one. 
In the present letter we re-examine the~dimensional reduction of $E_8$ over $SU(3)/ U(1) \times U(1) \times \mathbf{Z}_3$, where~the latter is the non-symmetric~coset space $SU(3)/U(1)\times U(1)$ equipped with~the freely acting discrete~symmetry $\mathbf{Z}_3$ in order that~the Wilson flux breaking~mechanism to get induced for further reduction~of the gauge symmetry of the $4D$ GUT, specifically~to $SU(3)^3$ along with two $U(1)$ global~symmetries \cite{Kapetanakis:1992hf,Manousselis:2001xb,Chatzistavrakidis:2008ii,Irges:2011de} (see also \cite{Lust:1985be}). The potential of the~resulting $4D$ theory contains terms that~can be identified as $F$-, $D$- and soft~breaking terms, which means that the~resulting theory is a (broken) $\mathcal{N}=1$ supersymmetric theory. 

In our case, the compactification~and unification scales coincide, leading~to a split-like supersymmetry scenario in which~some supersymmetric particles are superheavy, while others~obtain mass in the $TeV$ region. After the employment of~the spontaneous symmetry breaking of the GUT, the model~can be viewed as a two Higgs doublet~model (2HDM) which is phenomenologically consistent,~since it produces masses of~the light Higgs boson and the top and the bottom quarks within the experimental range.

\section{Dimensional Reduction of $E_8$ over $SU(3)/U(1)\times U(1)$}\label{drsu3u1x2z3}
In this section we focus directly on~the application of the CSDR scheme in~which we are interested. 
For a more complete~picture of the geometry of coset~spaces see \cite{Kapetanakis:1992hf,Castellani:1999fz}.
Also, for~the main aspects of the CSDR, the~generalized methodology of the reduction and the treatment~of the constraints,~see ref.\cite{Kapetanakis:1992hf}.

Let us now~demonstrate an illustrative example~of the CSDR scheme, that is the case of an $\mathcal{N}=1$ supersymmetric $E_8$ YM theory, which undergoes a dimensional~reduction over the non-symmetric coset space  $SU(3)/U(1)\times U(1)$ \cite{Kapetanakis:1992hf,Manousselis:2001xb,Lust:1985be}. The $4D$ YM action is:
\begin{align}
    S=C\int d^4x\,\mathrm{tr}\left[-\frac{1}{8}F_{\mu\nu}F^{\mu\nu}-\frac{1}{4}(D_\mu\phi_a)(D^\mu\phi^a)\right]+V(\phi)+\frac{i}{2}\bar{\psi}\Gamma^\mu D_\mu\psi-\frac{i}{2}\bar{\psi}\Gamma^aD_a\psi \,,
\end{align}
where it has~been identified:
\begin{align}
    V(\phi)&=-\frac{1}{8}g^{ac}g^{bd}\mathrm{tr}\left(f_{ab}^{~~C}\phi_C-ig[\phi_a,\phi_b])(f_{cd}^{~~D}\phi_D-ig[\phi_c,\phi_d]\right)\label{four-dimpotential}
\end{align}
and $\mathrm{tr}(T^iT^j)=2\delta^{ij}$, where $T^i$ are the generators of the~gauge group, $C$ is the volume of the~coset, $D_\mu=\partial_\mu-igA_\mu$ is the $4D$ covariant~derivative, $D_a$ is that of the coset and the coset metric is~given (in terms of its radii) by $g_{\alpha\beta}=\text{diag}(R_1^2,R_1^2,R_2^2,R_2^2,R_3^2,R_3^2)$.

The $4D$ gauge~group is determined~by the way~the $R=U(1)\times U(1)$ is embedded in $E_8$ and is obtained by~the centralizer of $R=U(1)\times U(1)$ in $G=E_8$, that~is:
\begin{align}
H=C_{E_8}(U(1)_A\times U(1)_B)=E_6\times U(1)_A\times U(1)_B\,.
\end{align}
Moreover, solving the constraints, the~scalar and fermion fields that remain in the $4D$ theory~are obtained by the decomposition of the~representation 248 -adjoint representation- of $E_8$ under $U(1)_A\times U(1)_B$. Also, in order to obtain the representations of the surviving fields of the $4D$ theory, it~is necessary to examine the decompositions of the vector~and spinor representations of $SO(6)$ under $R=U(1)_A\times U(1)_B$ (details in \cite{Kapetanakis:1992hf,Manousselis:2001xb,Irges:2011de}). Therefore, the CSDR rules imply~that the surviving gauge fields (those of $E_6\times U(1)_A\times U(1)_B$) are accommodated in three $\mathcal{N}=1$ vector
supermultiplets in the $4D$ theory.~Also, the matter fields of the $4D$ theory end up in six chiral~multiplets. Three of them are $E_6$ singlets~carrying $U(1)_A\times U(1)_B$ charges, while the
rest are chiral~multiplets. The unconstrained fields transforming~under $E_6\times
U(1)_A\times U(1)_B$ are:
\begin{align*}
\alpha_i \sim 27_{(3,\frac{1}{2})}, \quad  \beta_i \sim
27_{(-3,\frac{1}{2})}, \quad  \gamma_i \sim 27_{(0,-1)}, \quad
\alpha \sim 1_{(3,\frac{1}{2})}, \quad  \beta \sim
1_{(-3,\frac{1}{2})}, \quad  \gamma \sim 1_{(0,-1)}
\end{align*}
and the scalar potential of the theory is:
\begin{multline}
\frac{2}{g^2}V(\alpha^i,\alpha,\beta^i,\beta,\gamma^i,\gamma)= \frac{2}{5}\left(\frac{1}{R_1^4}+\frac{1}{R_2^4}+\frac{1}{R_3^4}\right)\\
+\bigg(\frac{4R_1^2}{R_2^2 R_3^2}-\frac{8}{R_1^2}\bigg)\alpha^i \alpha_i + \bigg(\frac{4R_1^2}{R_2^2 R_3^2}- \frac{8}{R_1^2}\bigg)\bar{\alpha} \alpha \\
+\bigg(\frac{4R_2^2}{R_1^2 R_3^2}-\frac{8}{R_2^2}\bigg)\beta^i \beta_i +\bigg(\frac{4R_2^2}{R_1^2 R_3^2}-\frac{8}{R_2^2}\bigg)\bar{\beta} \beta \\
+\bigg(\frac{4R_3^2}{R_1^2 R_2^2}-\frac{8}{R_3^2}\bigg)\gamma^i \gamma_i +\bigg(\frac{4R_3^2}{R_1^2 R_2^2}-\frac{8}{R_3^2}\bigg)\bar{\gamma} \gamma\\
+\bigg[\sqrt{2}80 \bigg(\frac{R_1}{R_2 R_3}+\frac{R_2}{R_1 R_3}+\frac{R_3}{R_2 R_1}\bigg)d_{ijk}\alpha^i \beta^j \gamma^k+\sqrt{2}80\bigg(\frac{R_1}{R_2 R_3}+\frac{R_2}{R_1 R_3}+\frac{R_3}{R_2 R_1}\bigg)\alpha \beta \gamma +h.c \bigg] \nonumber \\
+ \frac{1}{6}\bigg( \alpha^i(G^\alpha)_i^j\alpha_j+\beta^i(G^\alpha)_i^j\beta_j+\gamma^i(G^\alpha)_i^j\gamma_j\bigg)^2 \\
+\frac{10}{6}\bigg( \alpha^i(3\delta_i^j)\alpha_j+\bar{\alpha}(3)\alpha+\beta^i(-3\delta_i^j)\beta_j+\bar{\beta}(-3)\beta \bigg)^2 \\
+\frac{40}{6}\bigg( \alpha^i(\tfrac{1}{2}\delta_i^j)\alpha_j+\bar{\alpha}(\tfrac{1}{2})\alpha+\beta^i(\tfrac{1}{2}\delta_i^j)\beta_j+\bar{\beta}(\tfrac{1}{2})\beta+\gamma^i(-1\delta_i^j)\gamma^j+\bar{\gamma}(-1)\gamma \bigg)^2 \\
+40\alpha^i \beta^j d_{ijk}d^{klm} \alpha_l \beta_m+40\beta^i
\gamma^j d_{ijk}d^{klm} \beta_l
\gamma_m+40 \alpha^i \gamma^jd_{ijk} d^{klm} \alpha_l \gamma_m \\
+40(\bar{\alpha}\bar{\beta})(\alpha\beta)+40(\bar{\beta}\bar{\gamma})(\beta\gamma)+40(\bar{\gamma}\bar{\alpha})(\gamma
\alpha)\,,\label{deka}
\end{multline}
being also positive~definite. In the above expression of the scalar potential,~the $F-, D-$ and soft supersymmetry~breaking terms are~identified. The $F$-terms emerge from~the superpotential:
\begin{align}
\mathcal{W}(A^i,B^j,C^k,A,B,C)=\sqrt{40}d_{ijk}A^iB^jC^k+\sqrt{40}ABC\,,
\end{align}
where $A^i,B^j,C^k$ are the superfields that are accommodated in the $27$ representation and $A,B,C$ are the superfields that are singlets under $E_6$ and only have $U(1)$ charges.
The $D$-terms are structured~as:
\begin{align}
\frac{1}{2}D^{\alpha}D^{\alpha}+\frac{1}{2}D_1D_1+\frac{1}{2}D_2D_2\,,
\end{align}
where the $D$ quantities are~calculated as:
\begin{align*}
D^{\alpha}&=\frac{1}{\sqrt{3}}\Big(\alpha^i(G^{\alpha})_i^j\alpha_j+\beta^i(G^{\alpha})_i^j\beta_j+\gamma^i(G^{\alpha})_i^j\gamma_j\Big),\\
D_1&=\sqrt{\frac{10}{3}}\Big(\alpha^i(3\delta_i^j)\alpha_j+\bar{\alpha}(3)\alpha+\beta^i(-3\delta_i^j)\beta_j+\bar{\beta}(-3)\beta\Big)\\
D_2&=\sqrt{\tfrac{40}{3}}\Big(\alpha^i(\tfrac{1}{2}\delta_i^j)\alpha_j+\bar{\alpha}(\tfrac{1}{2})\alpha+\beta^i(\tfrac{1}{2}\delta_i^j)\beta_j+\bar{\beta}(\tfrac{1}{2})\beta+\gamma^i(-1\delta_i^j)\gamma_j+\bar{\gamma}(-1)\gamma\Big)\,.
\end{align*}
Apart from the~terms of the potential of \refeq{deka} identified~as $F$- and $D$- terms, the remaining ones admit the interpretation~of soft scalar masses and trilinear soft terms. The gaugino~demonstrates a special behaviour compared to the rest soft~supersymmetric terms, as can be seen in the following~relation\footnote{The relation shows that the gauginos naturally~obtain mass at the compactification scale \cite{Kapetanakis:1992hf}. This, however, is prevented by the inclusion of the~torsion \cite{Manousselis:2001xb}, which is the case in the following construction.}:
\begin{equation}
    M=(1+3\tau)\frac{R_1^2+R_2^2+R_3^2}{8\sqrt{R_1^2R_2^2R_3^2}}\,.
\end{equation}

The next step~is the minimization of~the potential, which requires at least two of the three~singlets to acquire vevs at the compactification scale. For the~purposes of our current work we choose~the singlets $\alpha$ and $\beta$ to acquire vevs, while $\gamma$ remains massless. This results in the breaking of the~two $U(1)$, reducing our gauge group from $E_6\times U(1)_A\times U(1)_B$ to $E_6$. The two abelian groups remain,~however, as global symmetries, which will be very useful in~conserving Baryon number, as it will be discussed~below.

\section{Breaking by Wilson Flux mechanism} \label{wilson}

In the above section we presented the~case in which the CSDR scheme is applied~on a higher-dimensional $E_8$ gauge theory which is reduced~over an $SU(3)/U(1)\times U(1)$ coset space and leads to~a $4D$ $E_6$ gauge theory. However, the $E_6$ group cannot be~broken exclusively by the presence of the $27$ Higgs multiplet. For~this reason, that is to reduce the~resulting gauge symmetry, the Wilson~flux breaking mechanism is~employed \cite{Kozimirov:1989kn,Zoupanos:1987wj,Hosotani:1983xw}. The below procedure can be found in detail in \cite{Irges:2011de}. 

\subsection{ $SU(3)^3$  produced by Wilson flux} \label{fluxsu33}
The Wilson~flux breaking mechanism, projects the theory in such a way that~the surviving fields are those which remain invariant under~the action of the freely acting discrete symmetry, the $\mathbf{Z}_3$, on their gauge~and geometric indices. The non-trivial~action of the $\mathbf{Z}_3$ group on the gauge indices of the~various fields is parametrized by the matrix \cite{Chatzistavrakidis:2010xi}:
\begin{equation}
    \gamma_3=\text{diag}\{\mathbf{1}_3,\omega \mathbf{1}_3, \omega^2
\mathbf{1}_3\}\,,
\end{equation}
where $\omega=e^{i\frac{2\pi}{3}}$. The latter~acts on the gauge fields of the $E_6$ gauge~theory and a non-trivial phase acts on the matter fields. First,~the gauge fields that pass through the filtering of the~projection are those which satisfy the condition:
\begin{equation}
[A_M,\gamma_3]=0\;\;\Rightarrow A_M=\gamma_3 A_M \gamma_3^{-1}\label{filteringgaugefields}
\end{equation}
and the remaining~gauge symmetry is $SU(3)_c\times SU(3)_L \times SU(3)_R.$                                        
The matter counterpart of \refeq{filteringgaugefields} is:
\begin{equation}
\vec{\alpha}=\omega\gamma_3\vec{\alpha},\;\;\vec{\beta}=\omega^2
\gamma_3\vec{\beta},\;\;\vec{\gamma}=\omega^3 \gamma_3\vec{\gamma}\,,\qquad \alpha=\omega \alpha,\;\;\beta=\omega^2 {\beta},\;\;{\gamma}=\omega^3 {\gamma}\,.
\end{equation}\label{wilsonprojectionmatter}
where $\vec \alpha,\vec \beta, \vec \gamma$ are the~matter superfields which belong to the 27
representation and $\alpha,\beta,\gamma$ the singlets~that only carry $U(1)_{A,B}$ charges. 
The representations of the remnant~group, $SU(3)_c\times SU(3)_L\times SU(3)_R$, in which~the above fields are accommodated, are obtained after~considering the decomposition rule of the 27 representation of $E_6$ under the new group, $(1,3,\bar 3)\oplus(\bar3,1,3)\oplus(3,\bar 3,1)$.
Therefore, in the projected theory we are left with the following matter content:
\begin{align*}
\alpha_3\equiv \Psi_1\sim (\bar{3},1,3)_{(3,\frac{1}{2})}, \;\;\; \beta_2\equiv
\Psi_2\sim (3,\bar{3},1)_{(-3,\frac{1}{2})}, \;\;\; \gamma_1\equiv \Psi_3\sim
(1,\bar{3},3)_{(0,-1)}, \;\;\; \gamma\equiv \theta_{(0,-1)},
\end{align*}
where the three former are the leftovers of $\vec \alpha,\vec \beta, \vec \gamma$ and together they form a 27 representation of $E_6$, that means that the leftover content can be identified as one generation. In order to obtain 
 a spectrum consisting of three generations, one may introduce non-trivial~monopole charges in the $U(1)$s in $R$, resulting in a total of three~replicas of the above~fields
(where an index $l = 1, 2, 3$ can be used to specify each of the three families).

The scalar potential of~the $E_6$ (plus the global abelian symmetries) that was~obtained after the dimensional~reduction of $E_8$, \refeq{deka}, can~now (that is after the adoption of the~Wilson flux breaking mechanism and the~projection of the theory) be rewritten~in the $SU(3)_c\times SU(3)_L\times SU(3)_R$ language as \cite{Irges:2011de}:
\begin{align}
V_{sc}=3\cdot \frac{2}{5}\Big(\frac{1}{R_1^4}+\frac{1}{R_2^4}+\frac{1}{R_3^4}\Big)+\underset{l=1,2,3}{\sum}V^{(l)}\,,
\end{align}
in which:
\begin{align}
V^{(l)}=V_{susy}+V_{soft}=V_D+V_F+V_{soft}\,.
\end{align}
From now on, we~give up on the generation superscript $(l)$, since our analysis will be focused on the third generation, and it will only be written explicitly when required. Regarding the $D$ and $F$-terms, they are identified as:
\begin{align}
V_D =\frac{1}{2}\underset{A}{\sum}D^AD^A+\frac{1}{2}D_1D_1+\frac{1}{2}D_2D_2, \\
V_F =\underset{i=1,2,3}{\sum}|F_{\Psi_i}|^2+|F_{\theta}|^2, \;\;
F_{\Psi_i}=\frac{\partial\mathcal{W}}{\partial \Psi_i} ,\;\;
F_{\theta}=\frac{\partial\mathcal{W}}{\partial \theta}\,,
\end{align}
where the $F$-terms derive~from the expression:
\begin{align}
\mathcal{W}=\sqrt{40}d_{abc}\Psi_1^a\Psi_2^b\Psi_3^c\,,
\end{align}
while the $D$-terms are written~explicitly as:
\begin{align}
D^A &=\frac{1}{\sqrt{3}}\big<\Psi_i|G^A|\Psi_i\big>, \label{DA} \\
D_1 &=3\sqrt{\frac{10}{3}}(\big< \Psi_1|\Psi_1\big>-\big<\Psi_2|\Psi_2\big>),\label{D1} \\
D_2 &=\sqrt{\frac{10}{3}}(\big<
\Psi_1|\Psi_1\big>+\big<\Psi_2|\Psi_2\big>-2\big<\Psi_3|\Psi_3\big>-2|\theta|^2)\,. \label{D2}
\end{align}
Last, the soft supersymmetry~breaking terms are written down as:
\begin{align}
V_{soft}=&\left(\frac{4R_1^2}{R_2^2R_3^2}-\frac{8}{R_1^2}\right)\big<\Psi_1|\Psi_1\big>+\left(\frac{4R_2^2}{R_1^2R_3^2}-\frac{8}{R_2^2}\right)\big<\Psi_2|\Psi_2\big>\nonumber \\
&+\left(\frac{4R_3^2}{R_1^2R_2^2}-\frac{8}{R_3^2}\right)(\big<\Psi_3|\Psi_3\big>+|\theta|^2)\nonumber \\
&+80\sqrt{2}\left(\frac{R_1}{R_2R_3}+\frac{R_2}{R_1R_3}+\frac{R_3}{R_1R_2}\right)(d_{abc}\Psi_1^a\Psi_2^b\Psi_3^c+h.c)\\
=& m_1^2\big<\Psi_1|\Psi_1\big>+m_2^2\big<\Psi_2|\Psi_2\big>+m_3^2\Big(\big<\Psi_3|\Psi_3\big>+|\theta|^2\Big)+(\alpha_{abc}\Psi_1^a\Psi_2^b\Psi_3^c+h.c)\,.
\end{align}
The $(G^A)_a^{~b}$ are the structure~constants of the $SU(3)_c\times SU(3)_L\times SU(3)_R$ and~therefore antisymmetric in $a$ and $b$. According~to ref.\cite{Kephart:1981gf}, the vectors of~the $27$ of $E_6$ can be written in a more convenient~form in the $SU(3)_c\times SU(3)_L\times SU(3)_R$ language,~that is in complex $3\times 3$ matrices. Identification of:
\begin{align}
\Psi_1\sim (\bar{3},1,3)\rightarrow (q^c)_p^{~\alpha} , \;\;\Psi_2\sim
(3,\bar{3},1)\rightarrow (Q^{~a}_{\alpha}), \;\; \Psi_3\sim
(1,3,\bar{3})\rightarrow L_a^{~p}\,,
\end{align}
leads to the following~relabeling and assignment of the particle content of the~MSSM (and more) in the above representation of the model:
\begin{eqnarray*}
  q^c=\left(\begin{array}{ccc}
 d^{c1}_R & u^{c1}_R & D^{c1}_R \\
 d^{c2}_R & u^{c2}_R & D^{c2}_R \\
 d^{c3}_R & u^{c3}_R & D^{c3}_R
 \end{array}
 \right)\,,\,\, Q=\left(\begin{array}{ccc}
 -d^1_L & -d^2_L & -d^3_L \\
 u^1_L & u^2_L & u^3_L \\
 D^1_L & D^2_L & D^3_L
 \end{array}\right)\,,\,\, L=\left(\begin{array}{ccc}
 H_d^0 & H_u^+ & \nu_L \\
 H_d^- & H_u^0 & e_L \\
 \nu^c_R & e^c_R & S
 \end{array}\right)\,.
\end{eqnarray*}
It is evident~from the above that $d_{L,R},u_{L,R},D_{L,R}$ transform as $3, \bar{3}$ under the
colour~group.
\section{Selection of parameters and GUT breaking}\label{GUTbreaking}
With the above-mentioned theoretical framework fully in place, it is time~to specify the compactification scale of the theory, as well as other (resulting) quantities, in order to proceed to phenomenology.

\subsection{Choice of radii}\label{radii}
We will examine the case where the compactification scale is high\footnote{In this case Kaluza-Klein excitations are irrelevant. Otherwise one would need the eigenvalues of the Dirac and Laplace operators in the $6D$ compactification space.}, and more specifically $M_C=M_{GUT}$. Thus for the radii we have
$R_l\sim \frac{1}{M_{GUT}}~, ~ l=1,2,3$.

Without any special treatment, this results in soft trilinear couplings and soft scalar masses around $M_{GUT}$. However, we can select our third radius slightly different than the other two in a way that yields:
\begin{equation}
m_3^2\sim-\mathcal{O}(TeV^2),~~~~m_{1,2}^2\sim-\mathcal{O}(M_{GUT}^2),~~~~a_{abc}\gtrsim M_{GUT}~.
\end{equation}
In other words, we have supermassive squarks and $TeV$-scaled sleptons. Thus, supersymmetry is softly broken already at the unification scale, in addition to its breaking by both $D$-terms and $F$-terms.

\subsection{Further gauge symmetry breaking of $SU(3)^3$}\label{breakings}
The spontaneous breaking of the $SU(3)_L$ and $SU(3)_R$ can be
triggered by the following vevs of the two families of $L$'s.
\[
\langle L_s^{(3)}\rangle=\left(\begin{array}{ccc}
0 & 0&0\\
0&0&0\\
0&0&V
\end{array}\right),\;\;
\langle L_s^{(2)}\rangle=\left(\begin{array}{ccc}
0 & 0&0\\
0&0&0\\
V&0&0
\end{array}\right)~,
\]
where the $s$ index denotes the scalar component of the multiplet. These vevs are singlets under $SU(3)_c$, so they leave the colour group unbroken. 
If we use only $\langle L_s^{(3)}\rangle$ we get the breaking
\begin{equation}
SU(3)_c\times SU(3)_L\times SU(3)_R \rightarrow SU(3)_c\times SU(2)_L\times SU(2)_R\times U(1)~,
\end{equation}
while if we use only $\langle L_s^{(2)}\rangle$ we get the breaking 
\begin{equation}
SU(3)_c\times SU(3)_L\times SU(3)_R \rightarrow SU(3)_c\times SU(2)_L\times SU(2)'_R\times U(1)'~.
\end{equation}
Their combination gives the desired breaking \cite{Babu:1985gi}:
\begin{equation}\label{desiredbreaking}
SU(3)_c\times SU(3)_L\times SU(3)_R \rightarrow SU(3)_c\times SU(2)_L\times U(1)_Y~.
\end{equation}
The configuration of the scalar potential just after the breaking gives vevs to the singlet of each family (not necessarily to all three). In our case we have $\langle\theta^{(3)}\rangle\sim\mathcal{O}(TeV)~,~\langle\theta^{(1,2)}\rangle\sim\mathcal{O}(M_{GUT})$.

\noindent Electroweak (EW) breaking then proceeds by the vevs
\cite{Ma:2004mi}:
\[
\langle L_s^{(3)}\rangle=\left(\begin{array}{ccc}
\upsilon_d& 0&0\\
0&\upsilon_u&0\\
0&0&0
\end{array}\right)\;.
\]

\subsection{Lepton Yukawa couplings and $\mu$ terms}\label{higherdimop}
Although the two $U(1)$s were already broken before the Wilson flux breaking, they still impose global symmetries. As a result, in the lepton sector we cannot have invariant Yukawa terms. However, below the unification scale, an effective term can occur from higher-dimensional operators \cite{Irges:2011de}:
\begin{equation}
L\overline{e}H_d\Big(\frac{\overline{K}}{M}\Big)^3~,
\end{equation}
where $\overline{K}$ is the vacuum expectation value of the  conjugate scalar component of either $S^{(i)},~\nu_R^{(i)}$ or $\theta^{(i)}$, or any combination of them, with or without mixing of flavours. Using similar arguments, one can also have mass terms for $S^{(i)}$ and $\nu_R^{(i)}$, which will then be rendered supermassive.

Another much needed quantity that is missing from our model is the $\mu$ term, one for each family of Higgs doublets. 
In the same way, we can have:
\begin{equation}
H_u^{(i)}H_d^{(i)}\overline{\theta}^{(i)}\frac{\overline{K}}{M}~.
\end{equation}
The first two generations of Higgs doublets will then have supermassive $\mu$ terms, while the $\mu$ term of the third generation will be at the $TeV$ scale.\\

\noindent In order to avoid confusion, it is useful to sum up the scale of some important parameters in Table 1.

\begin{center}
\begin{table}[ht]
\begin{center}
\small
\begin{tabular}{|l|r|}
\hline
 Parameter & Scale \\\hline
soft trilinear couplings & $\mathcal{O}(GUT)$  \\\hline
squark masses & $\mathcal{O}(GUT)$  \\\hline
slepton masses & $\mathcal{O}(TeV)$  \\\hline
$\mu^{(3)}$ & $\mathcal{O}(TeV) $  \\\hline
$\mu^{(1,2)}$ & $\mathcal{O}(GUT) $  \\\hline
unified gaugino mass $M_U$ & $\mathcal{O}(TeV) $  \\\hline
\end{tabular}
\caption{Approximate scale of parameters.}
\end{center}
\end{table}
\end{center}

\section{Phenomenological Analysis}\label{pheno}
Like every GUT, this model considers all gauge couplings to start as one coupling $g$ at $M_{GUT}$. However, since at the $E_8$ level there is only one coupling, it is clear that the (quark) Yukawa couplings are equal to $g$ at $M_{GUT}$ as well. This makes the selection of a large $tan\beta$ necessary. We use the unified coupling $g$ as a boundary condition for all the above-mentioned couplings at $M_{GUT}$. 

In our analysis we will use 1-loop beta functions for all parameters included. Below the unification scale they run according to the RGEs of the MSSM (squarks included) plus the 4 additional Higgs doublets (and their supersymmetric counterparts) that come from the two extra $L$ multiplets of the first and second generations, down to an intermediate scale $M_{int}$. Below this scale, all supermassive particles and parameters are considered decoupled, and the RGEs used include only the 2 Higgs doublets that originate from the third generation (and their respective Higgsinos), the sleptons and the gauginos. Finally, below a second intermediate scale that we call $M_{TeV}$, we run the RGEs of a non-supersymmetric 2HDM.

\subsection{Constraints}\label{constraints}
In our analysis we apply several experimental constraints, which we briefly review in this subsection.\\ 
Starting from the strong gauge coupling, we use the experimental value \cite{Zyla:2020zbs}:
\begin{equation}\label{astrong}
a_s(M_Z)=0.1187\pm0.0016\,.
\end{equation}
We calculate the top quark pole mass,~while the bottom quark~mass is evaluated~at $M_Z$, in order not to induce uncertainties that are inherent to~its pole~mass.  Their experimental values are \cite{Zyla:2020zbs}:
\beq
m_t^{\rm exp} = (172.4 \pm 0.7) \gev\,,~~~~~~ m_b(M_Z) = 2.83 \pm 0.10 \gev~.
\label{mtmbexp}
\eeq
We interpret the Higgs-like particle discovered in July 2012 by ATLAS and CMS
\cite{Aad:2012tfa}
as the light $\cal CP$-even Higgs~boson of~the supersymmetric SM.
The (SM) Higgs boson experimental average mass is \cite{Zyla:2020zbs}:
\beq
M_H^{\rm exp}=125.10\pm 0.14~{\rm GeV}~.\label{higgsexp}
\eeq

\subsection{Gauge unification}\label{unification}
A first challenge for each unification model is to predict a unification scale, while maintaining agreement with experimental constraints on gauge couplings. The 1-loop gauge $\beta$ fuctions are given by:
\begin{equation}
2\pi\beta_i=b_i\alpha_i^2~,
\end{equation}
where for the three energy regions the $b$ coefficients are given in Table 2.
 
\begin{center}
\begin{table}[ht]
\begin{center}
\small
\begin{tabular}{|l|r|r|r|}
\hline
 Scale & $b_1$ & $b_2$ & $b_3$ \\\hline
 $M_{EW}$-$M_{TeV}$ & $\frac{21}{5}$ & $-3$ & $-7$ \\\hline
 $M_{TeV}$-$M_{int}$ & $\frac{11}{2}$ & $-\frac{1}{2}$ & $-5$ \\\hline
 $M_{int}$-$M_{GUT}$ & $\frac{39}{5}$ & $3$ & $-3$ \\\hline
\end{tabular}
\caption{$b$ coefficients for gauge RGEs.}
\end{center}
\end{table}
\end{center}
The $a_{1,2}$ determine the unification scale and the $a_3$ is used to confirm that unification is indeed possible. Using a $0.3\%$ uncertainty at the unification scale boundary, we predict the different scales of our model (shown on Table 3), while the strong coupling is predicted within $2\sigma$ of the experimental value (\refeq{astrong}):
\begin{equation}
a_s(M_Z)=0.1218~.
\end{equation}
It should be noted that although the unification scale is somewhat lower than expected in a supersymmetric theory, there is no fear of fast proton decay, as the $U(1)_A$ remaining global symmetry can be immediately recognised as:
\begin{equation}
U(1)_A=-\frac{1}{9}B~,
\end{equation} 
where $B$ is the baryon number. Therefore, the unification scale could, in principle, lie even lower without such problems.
\begin{center}
\begin{table}[ht]
\begin{center}
\small
\begin{tabular}{|l|r|r|r|}
\hline
 Scale & $GeV$~~~~~  \\\hline
 $M_{GUT}$ & $\sim 1.7\times 10^{15}$  \\\hline
 $M_{int}$ & $\sim 9\times 10^{13}$~~ \\\hline
 $M_{TeV}$ & $\sim 1500$~~~~~~  \\\hline
\end{tabular}
\caption{Scale predicted by gauge unification.}
\end{center}
\end{table}
\end{center}

\subsection{Higgs potential}\label{2hdm}
We once again turn our focus on the third family. After GUT breaking, the Higgs scalar potential calculated from the $D$-, $F$- and soft terms of \refse{fluxsu33} is given by:
\begin{align}\label{higgspot}
V_{Higgs}=&\Big(3|\mu^{(3)}|^2+m_3^2\Big)\Big(|H_d^0|^2+|H_d^-|^2\Big)+\Big(3|\mu^{(3)}|^2+m_3^2\Big)\Big(|H_u^0|^2+|H_u^+|^2\Big)\nonumber\\
&+b^{(3)}\Big[(H_u^+H_D^--H_u^0H_D^0)+c.c.\Big]\nonumber\\
&+\frac{10}{3}g^2\Big[|H_d^0|^4+|H_d^-|^4+|H_u^0|^4+|H_u^+|^4+ \nonumber \\
&~~~~~~~~~~~~2|H_d^0|^2|H_d^-|^2+2|H_d^-|^2|H_u^0|^2+2|H_d^0|^2|H_u^+|^2+2|H_u^0|^2|H_u^+|^2\Big]\nonumber \\
&+\frac{20}{3}g^2\Big[|H_d^0|^2|H_u^0|^2+|H_d^-|^2|H_u^+|^2\Big]-20g^2\Big[\overline{H_d^0}H_d^-\overline{H_u^0}H_u^++c.c.\Big]~,
\end{align}
where $g$ is the gauge coupling at the unification scale and it is understood that the RG running has not yet taken place. One can easily compare the above potential with the standard 2 Higgs doublet scalar potential \cite{Gunion:1984yn,Quiros:1997vk,Branco:2011iw} and identify:
\begin{equation}
\lambda_1=\lambda_2=\lambda_3=\frac{20}{3}g^2~,~~~~~~\lambda_4=20g^2~,~~~~~~\lambda_5=\lambda_6=\lambda_7=0
\end{equation}\label{lambdas}
The above relations are used as boundary conditions at the GUT scale. Then, all the Higgs couplings run using their RGEs (see \cite{Haber:1993an} for the full expressions), which in turn change appropriately for each energy interval explained above.

\subsection{1-loop results}\label{1loopresults}
The Higgs couplings $\lambda_i$ are evolved from the GUT scale down to the EW scale together with the gauge couplings, the top, bottom and tau Yukawas, all at 1 loop. It is useful to remind the reader that all gauge and  quark Yukawa couplings use $g$ as boundary condition, while the tau Yukawa emerges from a higher-dimensional operator and has significantly wider freedom. We use the standard tau lepton mass \cite{Zyla:2020zbs} as an input. 

We consider uncertainties on the two important boundaries we consider, namely $M_{GUT}$ and $M_{TeV}$, because of threshold corrections (for a more comprehensive discussion see \cite{Kubo:1995cg}). For simplicity we have considered degeneracy between all supersymmetric particles that acquire masses at the $TeV$ scale.
The uncertainty of the top and bottom Yukawa couplings on the GUT boundary is taken to be $6\%$, while on the $TeV$ boundary is taken to be $2\%$. 
For $\lambda_{1,2}$ the uncertainty is $8\%$ on both boundaries and for $\lambda_{3,4}$ is $7\%$ at GUT and $5\%$ at $TeV$. 

Both top and bottom quark masses are predicted within $2\sigma$ of their experimental values (\refeq{mtmbexp}):
\begin{equation}
m_b(M_Z)=3.00~GeV~,~~~~~~~~~~\hat{m}_t=171.6~GeV~,
\end{equation}
while the light Higgs boson mass is predicted within $1\sigma$ of \refeq{higgsexp}:
\begin{equation}
m_h=125.18~GeV~.
\end{equation}
The model features a large $tan\beta\sim48$. This is necessary, since the Yukawas begin from the same value at the GUT boundary, so a large difference between the two vevs is needed to reproduce the known fermion hierarchy. The pseudoscalar Higgs boson is considered to have mass between $700-3000~GeV$.

The above 1-loop calculation could be subject to larger uncertainties, since they lack the precision of a higher-loop analysis.
The prediction of the full (light) supersymmetric spectrum, a 2-loop analysis of the model, the application of more experimental constraints (i.e. B-physics observables) and its discovery potential at present and/or future colliders are planned for future work \cite{future}.

\section{Conclusions}\label{conclusions}
Starting from~an $\mathcal{N}=1$, $10D$ $E_8$ Yang-Mills theory, we consider~a compactified~spacetime $M_4 \times B_0/ \mathbf{Z}_3 $, where $B_0$~is the non-symmetric~manifold
$SU(3)/U(1) \times U(1)$ and~$\mathbf{Z}_3$ is a~freely acting discrete~group on $B_0$.
Then we~reduce dimensionally the $E_8$ on~this manifold and we employ~the Wilson flux~mechanism
leading in~four dimensions to~an $\mathcal{N}=1~SU(3)^3$ gauge theory. We~consider the compactification~scale to match the~unification scale, a choice~that results in a split-like SUSY scenario, where gauginos, Higgsinos (of the third generation) and sleptons all acquire masses at $\sim1500~GeV$, and the rest supesymmetric spectrum is superheavy ($\sim M_{GUT}$). The global $U(1)_A$ conserves Baryon number, a fact which allows for the predicted unification scale $\sim10^{15}GeV$. The 2HDM employed below GUT predicts a light Higgs boson mass within the experimental limits, while the top and bottom quark masses are also in ($2\sigma$) agreement with experimental measurements.

We would like to thank our collaborators Sven Heinemeyer, Pantelis
Manousselis and Myriam Mondragon for their contribution in parts of the
present study and Peter Forgacs, Louis Ibanez, Dieter Lust, Stefan
Theisen, David Sutherland and Angel Uranga for constructive discussions.

\end{document}